\documentclass[%
  reprint,
superscriptaddress,
 amsmath,amssymb,
 aps,
prl,
nofootinbib
]{revtex4-2}

\usepackage{graphicx}
\usepackage{dcolumn}
\usepackage{bm}
\usepackage{hyperref}
\usepackage[dvipsnames]{xcolor}
\usepackage{orcidlink}
\usepackage{multirow}
\usepackage{aas_macros}
\usepackage{slashed}

\hypersetup{
  colorlinks   = true, 
  urlcolor     = Maroon, 
  linkcolor    = PineGreen, 
  citecolor   = Maroon 
}

\makeatletter
\def\maketitle{
\@author@finish
\title@column\titleblock@produce
\suppressfloats[t]}
\makeatother

\begin{document}


\title{Dimming Starlight with Dark Compact Objects}

\author{Joseph Bramante\orcidlink{0000-0001-8905-1960}}
\email{joseph.bramante@queensu.ca}
\affiliation{Department of Physics, Engineering Physics, and Astronomy, Queen’s University, Kingston, Ontario, K7L 3N6, Canada}
\affiliation{Arthur B. McDonald Canadian Astroparticle Physics Research Institute, Kingston ON K7L 3N6, Canada}
\affiliation{Perimeter Institute for Theoretical Physics, Waterloo, ON N2J 2W9, Canada}

\author{Melissa D. Diamond\orcidlink{0000-0003-1221-9475}}%
\email{m.diamond@queensu.ca}
\affiliation{Department of Physics, Engineering Physics, and Astronomy, Queen’s University, Kingston, Ontario, K7L 3N6, Canada}
\affiliation{Arthur B. McDonald Canadian Astroparticle Physics Research Institute, Kingston ON K7L 3N6, Canada}
\affiliation{Perimeter Institute for Theoretical Physics, Waterloo, ON N2J 2W9, Canada}
 
\author{J. Leo Kim\orcidlink{0000-0001-8699-834X}}%
\email{leo.kim@queensu.ca}
\affiliation{Department of Physics, Engineering Physics, and Astronomy, Queen’s University, Kingston, Ontario, K7L 3N6, Canada}
\affiliation{Arthur B. McDonald Canadian Astroparticle Physics Research Institute, Kingston ON K7L 3N6, Canada}

\begin{abstract}
We demonstrate a new technique to search for dark compact objects.  When dark matter comprising a dark compact object interacts with photons, the compact object can disperse light traveling though it.  As these objects pass between Earth and a distant star, they act as ``lampshades'' that dim the star.  We examine how dimming effects from clumps of dark matter in the Galaxy could be searched for in microlensing surveys, which measure the brightness of stars as a function of time. Using the EROS-2 and OGLE surveys, we show that a dimming analysis of existing data can be used to constrain dark sectors, and could be used to discover dark matter in compact objects. 

\end{abstract}

\maketitle

\newpage



\textbf{\textit{Introduction.}}---Primordial black holes (PBHs) \cite{Hawking:1971ei, Carr:1974nx} and MAssive Compact Halo Objects (MACHOs) are of great interest for beyond-Standard Model (BSM) studies of the Universe, both as dark matter (DM) candidates (see, $e.g.$, Refs.~\cite{Carr:2016drx, Green:2020jor, Villanueva-Domingo:2021spv}) and as natural consequences of many self-interacting dark sectors and exotic cosmologies \cite{Kouvaris:2015rea, Eby:2015hsq, Maselli:2017vfi, Amendola:2017xhl, Gresham:2017cvl, Chang:2018bgx, Savastano:2019zpr, Flores:2020drq, Domenech:2021uyx, Gurian:2021qhk, Hippert:2021fch, Flores:2022oef, Gurian:2022nbx, Domenech:2023afs, Flores:2023zpf, Roy:2023zar, Gemmell:2023trd, Bramante:2023ddr, Bramante:2024pyc, Roy:2024bcu}. There are several ways to observe and constrain these objects \cite{Giudice:2016zpa, VanTilburg:2018ykj, Cyr-Racine:2018htu, Dror:2019twh, Mishra-Sharma:2020ynk, Bai:2020jfm, Croon:2022tmr, Chen:2023xyj, Armstrong:2023cis, Banks:2023eym, Raj:2023azx, Tran:2023jci, Dent:2024yje, Kaplan:2024dsn}. Prominent among them are microlensing \cite{Paczynski1986} searches, which look for the amplification of starlight from a bright source due to gravitational lensing from an intermediate heavy object. Microlensing surveys, such as EROS-2 \cite{EROS-2:2006ryy}, OGLE \cite{ogle-iii, ogle-iv}, and Subaru/HSC \cite{Niikura:2017zjd}, have played an essential role in the hunt for BSM physics -- providing some of the most sensitive probes of PBHs and MACHOs in the $10^{-10} M_{\odot}$ to $10 M_{\odot}$ mass range \cite{Carr_2021}.

While microlensing searches traditionally focus on extremely compact  and heavy objects,   recent works \cite{Croon:2020wpr,Croon:2020ouk}, (and sources contained in Ref.~\cite{Croon:2024jhd}) have demonstrated that microlensing constraints change significantly for extended DM lenses. Further, these surveys may be sensitive to objects too light or diffuse to act as traditional lenses.  Dark compact objects with weak interactions with the Standard Model (SM) can scatter starlight.  When they pass between Earth and a distant source star this scattering dims the incoming starlight like a ``lampshade''. Microlensing surveys, which are sensitive to slight changes in the brightness of distant stars as a function of time, are well equipped to look for this dimming. Excitingly, these dimming effects are most observable for objects microlensing surveys are otherwise insensitive to, broadening their reach. 

In this Letter, we study how diffuse, self-gravitating clouds of DM which interact with SM photons dim starlight. The idea is similar to Ref. \cite{Bai:2023mfi}, which studied how dark exoplanets  gravitationally bound to a source star can dim the source starlight  and be observed in exoplanet searches. Here we consider free-floating dark structures in the Galaxy, which can enter the observing tube between the source star and the local observer. In particular, we adapt the existing microlensing events formalism \cite{Griest:1990vu, MACHO:1995udp} for  dimming events by defining a dimming event as a drop in apparent magnitude, and by replacing the Einstein radius of the object with its physical radius. 

\begin{figure}
\includegraphics[width=\columnwidth]{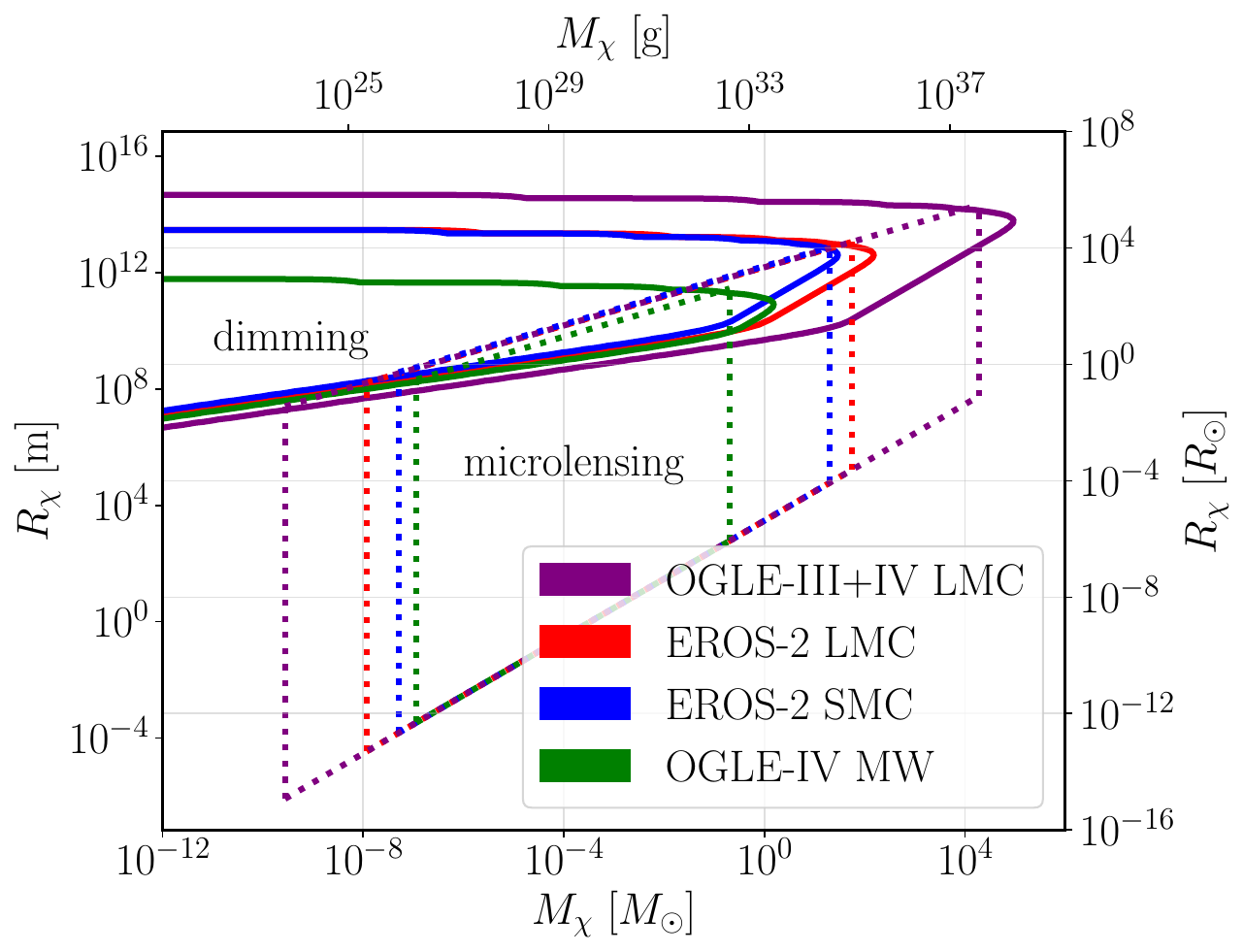}
\caption{\label{fig:heuristic_params} Heuristic plot showing various microlensing surveys' sensitivity to DM clumps assuming that $R_{\chi, \mathrm{eff}} \approx R_\chi$. Surveys are sensitive to dimming in the region enclosed by the solid curves, whereas surveys are sensitive to microlensing inside the dashed lines  \cite{Croon:2020wpr}.}
\end{figure}

We show various microlensing surveys' sensitivities to dimming effects, which are summarized in Fig. \ref{fig:heuristic_params}. As the amount of dimming depends on the SM photon-DM interaction, we first show how dimming effects could constrain a general mass and photon scattering cross section of the DM particles that make up the compact structures. We then show how dimming effects could be used to place complementary constraints on a specific DM model, using millicharged DM \cite{Holdom:1985ag} as an example. We show that null observations of dimming events may place competitive bounds on millicharged particles confined in compact objects. We remain agnostic about how these dark compact objects form, but note that many mechanisms for MACHO formation in the early Universe have been identified recently  \cite{Kouvaris:2015rea, Eby:2015hsq, Maselli:2017vfi, Amendola:2017xhl, Gresham:2017cvl, Chang:2018bgx, Savastano:2019zpr, Flores:2020drq, Domenech:2021uyx, Gurian:2021qhk, Hippert:2021fch, Flores:2022oef, Gurian:2022nbx, Domenech:2023afs, Flores:2023zpf, Roy:2023zar, Gemmell:2023trd, Bramante:2023ddr, Bramante:2024pyc, Roy:2024bcu}.


\textbf{\textit{Dimming from dark matter lampshades}}---To determine stellar dimming by DM objects, we treat the source star as point-like and consider an extended spherical DM cloud with constant density $\rho_\chi$, unless otherwise specified, and physical radius $R_\chi$. The transmittance ($i.e.$ the amount of starlight that passes through the DM cloud) at a position a distance $r$ from the center of the cloud is given by 
\begin{align}
    T(r) = \exp \left( - 2 \tau_0 \sqrt{1 - (r/R_\chi)^2 } \right), \label{eq:T_r}
\end{align}
where $\tau_0 \equiv R_\chi n_\chi \sigma$ is the characteristic optical depth of the DM cloud. The cross section $\sigma$ describes the interactions between DM particles and photons that cause dimming, and $n_\chi = \rho_\chi/m_\chi$ is the number density of the $\chi$ particles in the DM cloud. 
To trigger a dimming event, we require a minimum fraction $\mu_0$ of light be blocked by the cloud.  This occurs at an effective radius 
\begin{align}
    R_{\chi,\mathrm{eff}} = R_\chi \sqrt{1 - \frac{1}{4\tau_0^2} [\ln (1 - \mu_0)]^2  . \label{eq:R_eff}
}
\end{align}
The multiplicative factor on the right-hand side of Eq.~\eqref{eq:R_eff} is analogous to the usual impact parameter $u_T$ in the case of microlensing \cite{Griest:1990vu}. Here, we consider the dimming from a compact object transit to be detectable if the background source passes within $R_{\chi,\mathrm{eff}}$ of the center of the DM sphere. Note that Eq.~\eqref{eq:R_eff} gives a minimum value of $\tau_0$ to get a real solution, which is given by $\tau_{0, \mathrm{min}} = \frac{1}{2} \ln (1-\mu_0)$. Therefore, if a dimming event occurs, the measured flux of the source star is diluted to a fraction $1 - \mu_0 = 10^{( \bar{m} - m_\mathrm{dim} )  /2.5 }$, where $m_\mathrm{dim}$ is the dimmed apparent magnitude and $\bar{m}$ is the fiducial apparent magnitude of the source. 

For a given value of the characteristic optical depth $\tau_0$, the size of the DM cloud can be determined by the dip in magnitude. Microlensing surveys set a minimum total amplification to register a lensing event, given conventionally by $\mu_T = 1.34$ \cite{Paczynski1986} -- $i.e.$ a 34\% amplification. No convention exists for dimming, so we choose $\mu_0 = 0.34$ as the fiducial value throughout this Letter so that a dimming event occurs when 34\% of the starlight is dimmed. If a fiducial value of $\mu_0 = 0.34$ is selected, then $R_{\chi, \mathrm{eff}} \approx R_\chi $ for $\tau_0 \geq 1$. Hereafter we work with $\tau_0 = 1$ unless stated otherwise.

Especially for lower dimming thresholds ($i.e.$ 1\%), it will be important to consider other astrophysical sources of star dimming, $e.g.$ exoplanets, which are an active topic of research. Assuming 0.1 Jupiter-size exoplanets per star equally distributed from 1-100 AU \cite{Cumming_2008}, we estimate 
$\mathcal{O}$(100) 1\% dimming events in an OGLE-like 20 year dataset (removing planets that would be identifiable by orbiting repeatedly during the survey time). To differentiate between background dimming events and those from DM clumps, a careful light curve analysis could be employed, which is beyond the scope of this work. Prior work in Ref.~\cite{Bai:2023mfi} has shown that mock light curves from dark exoplanets and regular exoplanets are differentiable so long as the transiting object is sufficiently large and translucent. In the future it may be fruitful to conduct a similar study for microlensing-style surveys. 

\textbf{\textit{Dimming event rates.}}---We model the dimming event rates by modifying the procedure used for microlensing \cite{Croon:2020wpr, Croon:2020ouk}. We assume all DM clumps have a mass $M_\chi$, and velocities set by a Maxwell-Boltzmann distribution. We assume that the line-of-sight density of clumps in the Galaxy is given by $\rho_\mathrm{clumps}(x) = f_\mathrm{DM} \rho_\mathrm{DM}(x)$, where $f_\mathrm{DM}$ is the fraction of DM residing in these clumps and $\rho_\mathrm{DM}$ is the position-dependent DM density. Note that we have used the conventional variable $x = D_L/D_S$ to parameterize distances, where $D_L$ is the distance to the clump (normally the gravitational lens in microlensing scenarios), and $D_S$ is the distance to the source field. The DM density for the Milky Way (MW) is modelled assuming an isothermal profile
\begin{align}
    \rho_\mathrm{DM} (r(x)) = \frac{\rho_s}{1 + (r(x)/r_s)^2},
\end{align}
where $r_s = 4.38$ kpc is the central scale radius, and
\begin{align}
    r(x) = \sqrt{ R_\mathrm{Sol}^2 - 2 x R_\mathrm{Sol} D_S \cos \ell \cos b + x^2 D_S^2 },
\end{align}
in which $R_\mathrm{Sol} = 8.5$ kpc is the distance to the sun and $\rho_s = 1.39 $ GeV/cm$^{3}$ is the central density \cite{Cirelli:2010xx}. The differential event rate for microlensing events is given by \cite{Griest:1990vu, MACHO:1995udp}
\begin{align}
    \frac{d^2 \Gamma}{dxdt_E} = \varepsilon(t_E) \frac{2D_S}{v_0^2M_\chi} f_\mathrm{DM} \rho_\mathrm{DM}(x)v_E^4(x)e^{-v_E^2(x)/v_0^2}, \label{eq:differential_events}
\end{align}
where $v_0 = 220$ km/s is the circular speed of DM in our galaxy, $v_E$ is the relative velocity of the lensing system for the duration of the event, and $\varepsilon(t_E)$ is the so-called efficiency parameter. We take a constant value of $\varepsilon(t_E) = 1$, as evaluating the efficiency parameter for dimming is outside the scope of this study. Given this rate, the resulting number of events is
\begin{align}
    N_\mathrm{events} = N_* T_\mathrm{obs} \int_0^1 dx \int_{t_{E, \mathrm{min}}}^{t_{E, \mathrm{max}}} dt_E \frac{d^2 \Gamma}{dx dt_E} \label{eq:N_events} ,
\end{align}
where $t_{E,\mathrm{min}}$ ($t_{E,\mathrm{max}}$) is the minimum (maximum) timescale that the survey is sensitive to, $N_*$ is the number of stars in the source field, and $T_\mathrm{obs}$ is the total observing time of the survey. 

\begin{table*}[]
\centering
\begin{tabular}{cccccc}
 Survey  & Source & ($D_S$ [kpc], $\ell$, $b$) & $N_*$ & $T_\mathrm{obs}$ [days] & $t_E$ [days] \\ \hline\hline
 \multirow{2}{*}{EROS-2 \cite{EROS-2:2006ryy}}  & LMC & ($50, 280.46^\circ, -32.89^\circ $) & $5.49 \times 10^6$ & \multirow{2}{*}{2500} & \multirow{2}{*}{$[1,1000]$} \\ 
  & SMC & ($60, 302.81^\circ, -44.33^\circ $) & $0.86 \times 10^6$ &  &  \\ \hline
 OGLE-IV \cite{Niikura:2019kqi} & MW & ($8.5, 1.09^\circ, -2.39^\circ $) & $4.88 \times 10^7$  & 1826 & $[0.1,300]$ \\ \hline
 OGLE-III+IV \cite{Mroz:2024mse, Mroz:2024wag}  & LMC & ($50, 280.46^\circ, -32.89^\circ $) & $7.87 \times 10^7$ & 7300 & $[1,7300]$
\end{tabular}
\caption{Table of characteristic information of microlensing surveys used for Fig. \ref{fig:heuristic_params}, modified from Ref. \cite{Croon:2020wpr} which used data from Ref. \cite{Wenger:2000sw}, with additional data pertaining to OGLE-III+IV \cite{Mroz:2024mse, Mroz:2024wag}.}
\label{tab:surveys}
\end{table*}

Next we note that $t_{E,\mathrm{min}}$ and $t_{E,\mathrm{max}}$ are  determined by the behaviour of their respective efficiency parameters, which are related to the observing cadence and total observation time of the detector. Here we will use the minimum and maximum event times computed by the microlensing surveys \cite{EROS-2:2006ryy, Niikura:2019kqi, Mroz:2024mse, Mroz:2024wag}, except for OGLE-III+IV, where we have taken $t_{E,\mathrm{max}} = 7300$ days (corresponding to 20 years) rather than the 10000 days the analysis was performed up to. The microlensing event rate and count can easily be recast for dimming effects. The primary difference comes from considering the physical radii of the clumps rather than their Einstein radii \cite{Croon:2020wpr, Croon:2020ouk}. In our case, $v_E = 2R_{\chi, \mathrm{eff}}/t_E. $

We consider four source fields in our analysis: the EROS-2 surveys of the Large Magellanic Cloud (LMC) and the Small Magellanic Cloud (SMC) \cite{EROS-2:2006ryy}, the 5-year dataset from the OGLE-IV survey of the Milky Way bulge \cite{Niikura:2019kqi}, and the 20-year dataset from the combined OGLE-III \cite{ogle-iii} and OGLE-IV \cite{ogle-iv} (henceforth OGLE-III+IV) surveys of the LMC \cite{Mroz:2024mse, Mroz:2024wag}. The survey parameters needed to estimate the event rate are summarized in Table \ref{tab:surveys}. We have taken the relevant values for EROS-2 and OGLE-IV from Ref.~\cite{Croon:2020wpr} and adopted parameters from Refs.~\cite{Mroz:2024mse, Mroz:2024wag} for OGLE-III+IV. For simplicity, we considered the same range of event times for dimming events as previous microlensing studies. Note that while there is an 8-year dataset for the OGLE-IV survey of the Milky Way bulge \cite{ogle-iv-MW}, we have used the 5-year dataset for comparison with Ref.~\cite{Croon:2020wpr} and also since any constraints from the 8-year dataset will be weaker than the 20-year OGLE-III+IV dataset of the LMC.

As already mentioned, Fig. \ref{fig:heuristic_params}, which can be compared to microlensing mass reach figures in Refs.~\cite{Croon:2020wpr, Croon:2020ouk}, shows a heuristic estimate of the masses $M_\chi$ and physical radii $R_\chi$ of spherical DM clumps accessible to EROS-2, OGLE-IV, and OGLE-III+IV microlensing surveys. We do not assume a specific density profile in this figure.  To demonstrate the sensitivity of a survey to dimming and microlensing events, all enclosed regions have at least one event following Eq.~\eqref{eq:N_events}. The regions enclosed by the solid curves are where the surveys are sensitive to dimming effects, while the regions enclosed by dotted lines are where the surveys would be sensitive to microlensing. Note that the region for sensitivity for microlensing presented in Fig. \ref{fig:heuristic_params} is different from that shown in Ref.~\cite{Croon:2020wpr}, since we are considering a detection efficiency of $\varepsilon(t_E) = 1$ for both microlensing and dimming to present a fair comparison between ``best-case'' scenarios. We have also used full observation event time ranges for microlensing presented in Table \ref{tab:surveys} instead of a range where the efficiency parameter was roughly constant, as in Ref.~\cite{Croon:2020wpr}.

While microlensing events are generally only observable when the physical radius of the lens is smaller than its Einstein radius, dimming can occur in objects that are much larger and more diffuse. Dimming allows for detection not only of larger clumps than microlensing does, between $10^{-2} R_\odot$ and $10^6 R_\odot$ depending on the survey, but also lower mass clumps, since dimming depends on the density profile instead of the clump mass.

While Fig. \ref{fig:heuristic_params} implies dimming sensitivity for arbitrarily small clump masses, in practice there will be a lower cutoff in $M_\chi$ for the sensitivity region. This will occur when the clump does not have enough material to observably dim starlight, which will depend on the density profile of the object and the chosen interaction with the SM photon. Without specifying an explicit relationship for a density profile between $M_\chi$ and $R_\chi$, the number of events, $N_\mathrm{events}$, will continue to increase as $M_\chi$ decreases due to there being more clumps $c.f.$ Eq. \eqref{eq:differential_events}.  Note that in Fig.~\ref{fig:heuristic_params}, we assume that the effective dimming radius is equal to the clump radius. $R_{\chi, \mathrm{eff}} \approx R_\chi$, to demonstrate the best-case scenario for sensitivity to dimming effects. As OGLE-III+IV has the largest region of sensitivity, for the remainder of this Letter we focus only on OGLE-III+IV.

\begin{figure}
\includegraphics[width=\columnwidth]{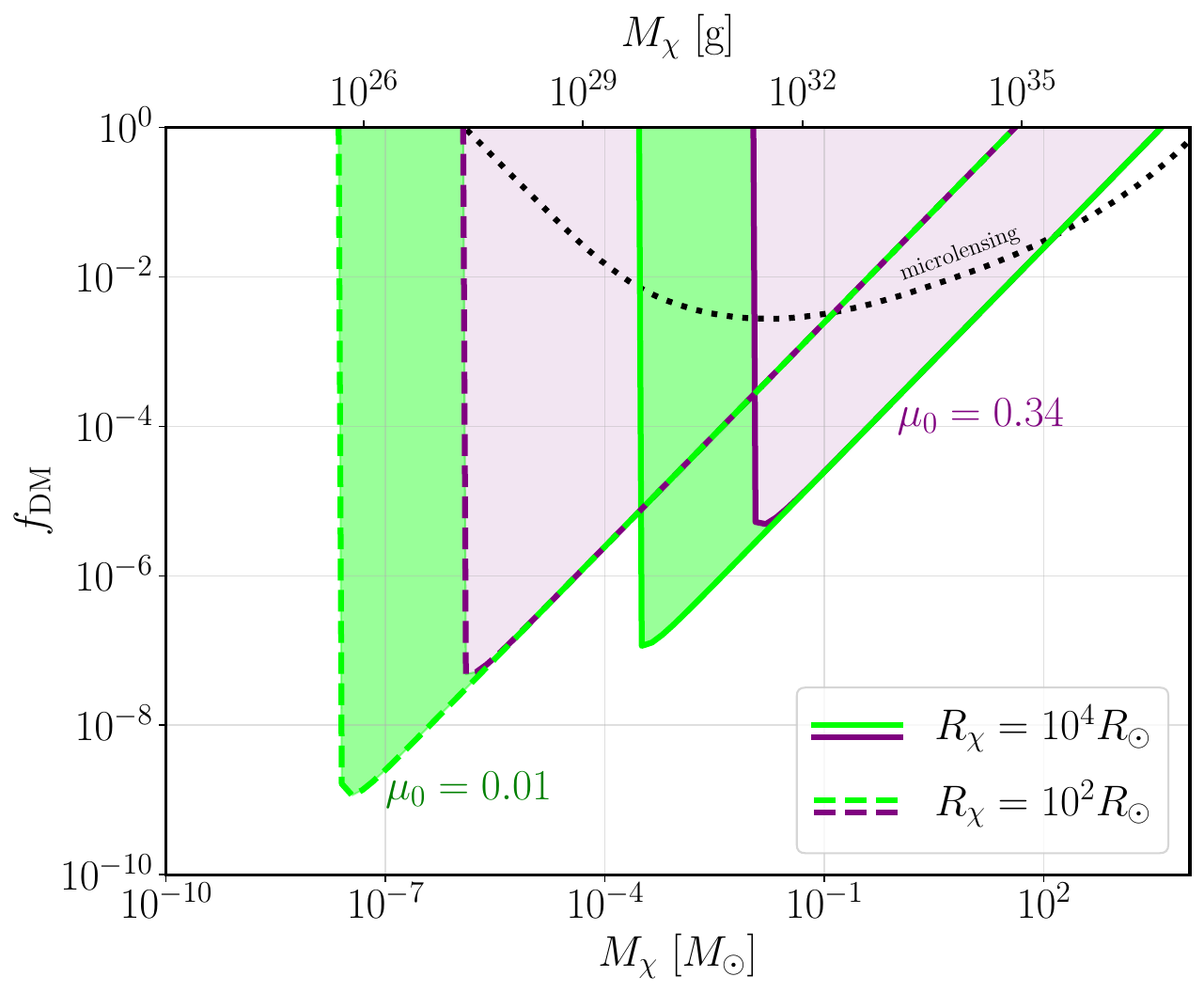}
\caption{\label{fig:f_example_ogle_fixed_R} Limits on the fraction of DM $f_\mathrm{DM}$ inside clumps of mass $M_\chi$ and fixed size $R_\chi = 10^4 R_\odot$ in solid curves and $R_\chi = 10^2 R_\odot$ in dashed curves, assuming null observations of dimming events in OGLE-III+IV. The regions enclosed and shaded in purple assume a dimming of $\mu_0 = 0.34$, while the lime green regions assume a dimming of $\mu_0 = 0.01$. We have fixed $\sigma/m_\chi = 10^2$ GeV$^{-3}$. The dotted curves in black show restrictions on $f_\mathrm{DM}$ assuming microlensing from point sources from OGLE-III+IV \cite{Mroz:2024mse, Mroz:2024wag}.}
\end{figure}

For a clump of constant density $M_\chi = 4\pi \rho_\chi R_\chi^3/3 $. Its radius is then given by
\begin{align}
    R_\chi = \sqrt{\frac{3}{4\pi} \frac{\sigma M_\chi }{m_\chi \tau_0} }.\label{eq:R_chi}
\end{align}
With a set density profile, $N_\mathrm{events}$ no longer monotonically decreases as a function of $M_\chi$, and so a minimum mass can be obtained. One can fix $R_\chi$ and solve for $\tau_0$ using Eq.~\eqref{eq:R_chi}, then compute the effective radius $R_{\chi, \mathrm{eff}}$ in Eq.~\eqref{eq:R_eff} if $\tau_0 > \tau_{0,\mathrm{min}}$ so that $R_{\chi, \mathrm{min}}$ is real and non-zero. Fig. \ref{fig:f_example_ogle_fixed_R} demonstrates how constraints on $f_\mathrm{DM}$ would look for clumps of fixed physical size. To compute these limits on $f_\mathrm{DM}$, we have taken $N_\mathrm{thresh} = 3.9$ events as the threshold number of events required for a 90\% confidence level observation of a single event, following Poisson statistics \cite{Croon:2020wpr}. The regions shaded above the curves indicate $N_\mathrm{events} \geq N_\mathrm{thresh}$, where null observations of dimming events in OGLE-III+IV could be used to set constraints for a fraction of DM, $f_\mathrm{DM} $, at a given DM clump mass.  The constraints cut off suddenly for low values of $M_\chi$ due to the optical depth's dependence on clump mass, $\tau_0 \propto M_\chi$, given in Eq.~\eqref{eq:R_chi}. When $M_\chi$ is large enough the clumps become opaque and $R_{\chi, \mathrm{eff}} \approx R_\chi$. Conversely, decreasing $M_\chi$ drives down the implied optical depth $\tau_0$.  As is apparent in  Eq.~\eqref{eq:R_eff} the effective radius quickly becomes zero once the required optical depth $\tau_0$ becomes too large.  The low mass cutoff occurs for $\tau_0 \approx \tau_{0, \mathrm{min}}$, where $M_{\chi, \mathrm{min}} = 2 \pi m_\chi R_\chi^2 \ln(1-\mu_0) /(3\sigma)$, so the minimum observable mass is determined by the minimum dimming required for an event observation.


\begin{figure}
\includegraphics[width=\columnwidth]{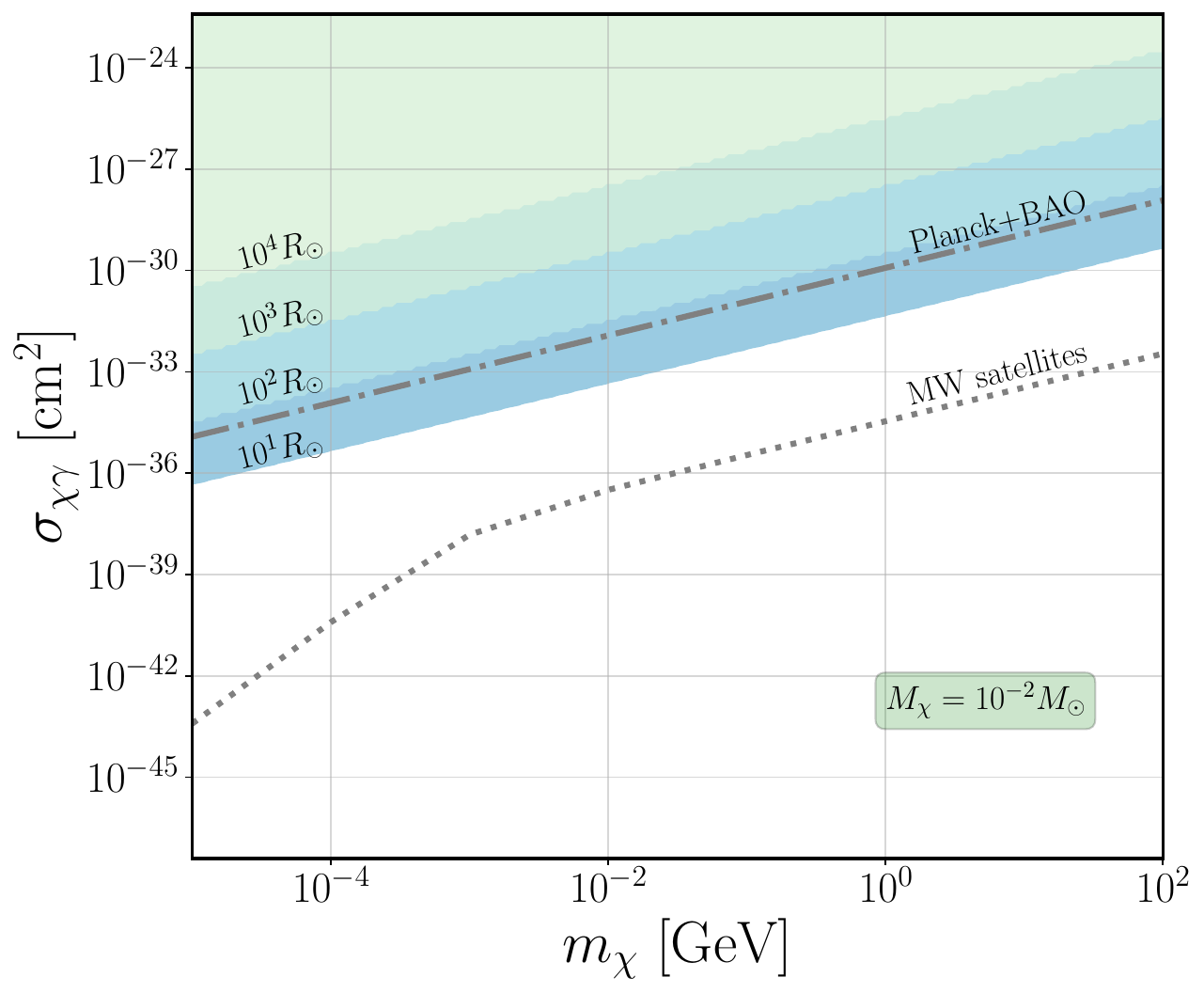}
\caption{\label{fig:consts_sigma} Possible constraints on $m_\chi$ and $\sigma_{\chi \gamma}$ for elastic scattering of SM photons and $\chi$-particles from a future null analysis of dimming events in the OGLE-III+IV surveys, assuming fixed physical sizes $R_\chi$ between $10$ to $10^4 R_\odot$, and a fixed mass of $M_\chi = 10^{-2} M_\odot$. We have assumed that these objects make up $f_\mathrm{DM} = 0.1$ of the DM.  The overlaid gray dashed-dotted lines are bounds on the elastic scattering cross section from Planck+BAO \cite{Becker:2020hzj}, while the gray dotted line are bounds from MW satellite galaxies \cite{Crumrine:2024sdn}. Note that these constraints are calculated for free floating DM with $f_{\text{DM}}=1$  and may not be applicable to the scenario shown in this plot.}
\end{figure}

\textbf{\textit{Dark matter--photon elastic scattering.}}---While previously we related the number of expected dimming events to the macroscopic properties of the DM clumps, we now show how dimming effects can be used to constrain the microscopic properties of DM comprising the clumps. This is explicitly detailed for elastic DM-SM photon scattering and for millicharged DM. By considering dimming due to elastic scattering $\chi \gamma \to \chi \gamma$ between DM particles and SM photons, Fig.~\ref{fig:consts_sigma} shows how the null observation of dimming events in OGLE-III+IV would constrain $m_\chi$--$\sigma_{\chi \gamma}$ parameter space, assuming various fixed physical radii for clump masses of $M_\chi = 10^{-2} M_\odot$. The high-$m_\chi$, low-$\sigma_{\chi \gamma}$ behaviour of the constraints is due to the DM objects becoming too small for detection. Note that at a fixed clump radius, more massive objects are generally more constrained, which is discussed further in the Supplemental Material. However, since $N_\mathrm{events} \propto M_\chi^{-1}$ in the high clump mass limit, these constraints eventually vanish below $N_\mathrm{thresh}$.

We have also plotted constraints on $\sigma_{\chi \gamma}$ from analyses of Planck and baryonic acoustic oscillation (BAO) data \cite{Becker:2020hzj}, as well as constraints from considering the population of MW satellites \cite{Crumrine:2024sdn}. However, we note that these constraints have assumed that all of DM is able to interact with the SM photon. While they are expected to weaken as one decreases the fraction of DM which is able to interact with the SM photon, their exact behavior under these circumstances remains unknown.   The constraints described in Ref.~\cite{Crumrine:2024sdn} that rely on DM photon-scattering which wash out small scale DM structures may behave quite differently than expected in Ref.~\cite{Crumrine:2024sdn} in the case where all of the $\chi$ reside in compact objects. Therefore when the $\chi$ particles responsible for dimming are a subdominant component of DM, dimming effects could provide competitive bounds for $m_\chi > 10^{-3}$ GeV.

\textbf{\textit{Millicharged dark matter.}}---Next we demonstrate how the lampshade effect can used to search for millicharged DM. Here we assume a simple asymmetric dark sector, containing a dark Dirac fermion $\chi$ with mass $m_\chi$ and a light dark photon $V_\mu$, which kinetically mixes with the SM photon $A_\mu$,
\begin{align}
    \mathcal{L} \supset  & \bar{\chi} (i \slashed{D} - m_\chi) \chi  - \frac{1}{4} V_{\mu \nu} V^{\mu \nu}
     - \frac{\epsilon}{2}V_{\mu\nu} F^{\mu\nu} , \label{eq:Lagrangian}
\end{align}
where $F_{\mu\nu} = \partial_\mu A_\nu - \partial_\nu A_\mu$ is the usual field strength tensor for the SM photon and $V_{\mu \nu}$ is similarly defined for the dark photon, so the last term in Eq.~\eqref{eq:Lagrangian} is a kinetic mixing term with a kinetic mixing parameter $\epsilon$. $D_\mu = \partial_\mu - i g_D V_\mu$ is the covariant derivative with $\alpha_D = g_D^2/(4\pi)$. We assume the dark photon obtains a mass from a Stueckelberg field charged under both the dark and SM Abelian gauge groups, which implies that $\chi$ will be electromagnetically millicharged \cite{Feldman:2007wj,Kahlhoefer:2020rmg}. For simplicity we fix parameters so that $\chi$'s have effective SM electric charge $q_\chi = \epsilon g_D / e $, where $e$ is the SM electromagnetic coupling. We note that for a sufficiently light dark photon, present bounds permit the mixing parameter $\epsilon$ to be large, $e.g.$ \cite{ParticleDataGroup:2024cfk,Caputo:2021eaa}. This class of models \cite{Holdom:1985ag} has been widely studied both as a BSM particle and a DM candidate (see, $e.g.$ \cite{Kelly:2018brz, Plestid:2020kdm, ArguellesDelgado:2021lek, Berlin:2022hmt, Gan:2023jbs, Wu:2024iqm, Berlin:2024lwe}).

SM photons passing through a dark clump can undergo various interactions with $\chi$ particles. Both scattering, $\chi \gamma  \to \chi\gamma $, and conversion to dark photons, $ \chi \gamma \to \chi \gamma'$, will cause apparent dimming. The scattering cross section is of order $\mathcal{O}(\epsilon^4)$, while the conversion cross section is of order $\mathcal{O}(\epsilon^2)$, making it the dominant process for dimming in our analysis. In the low energy limit, the $ \chi \gamma \to \chi \gamma'$ cross section is given by the Thomson cross section multiplied by a kinetic mixing parameter,
$ 
    \sigma = 8 \pi \alpha_D^2 \epsilon^2/(3 m_\chi^2).
$ 

\begin{figure}
\includegraphics[width=\columnwidth]{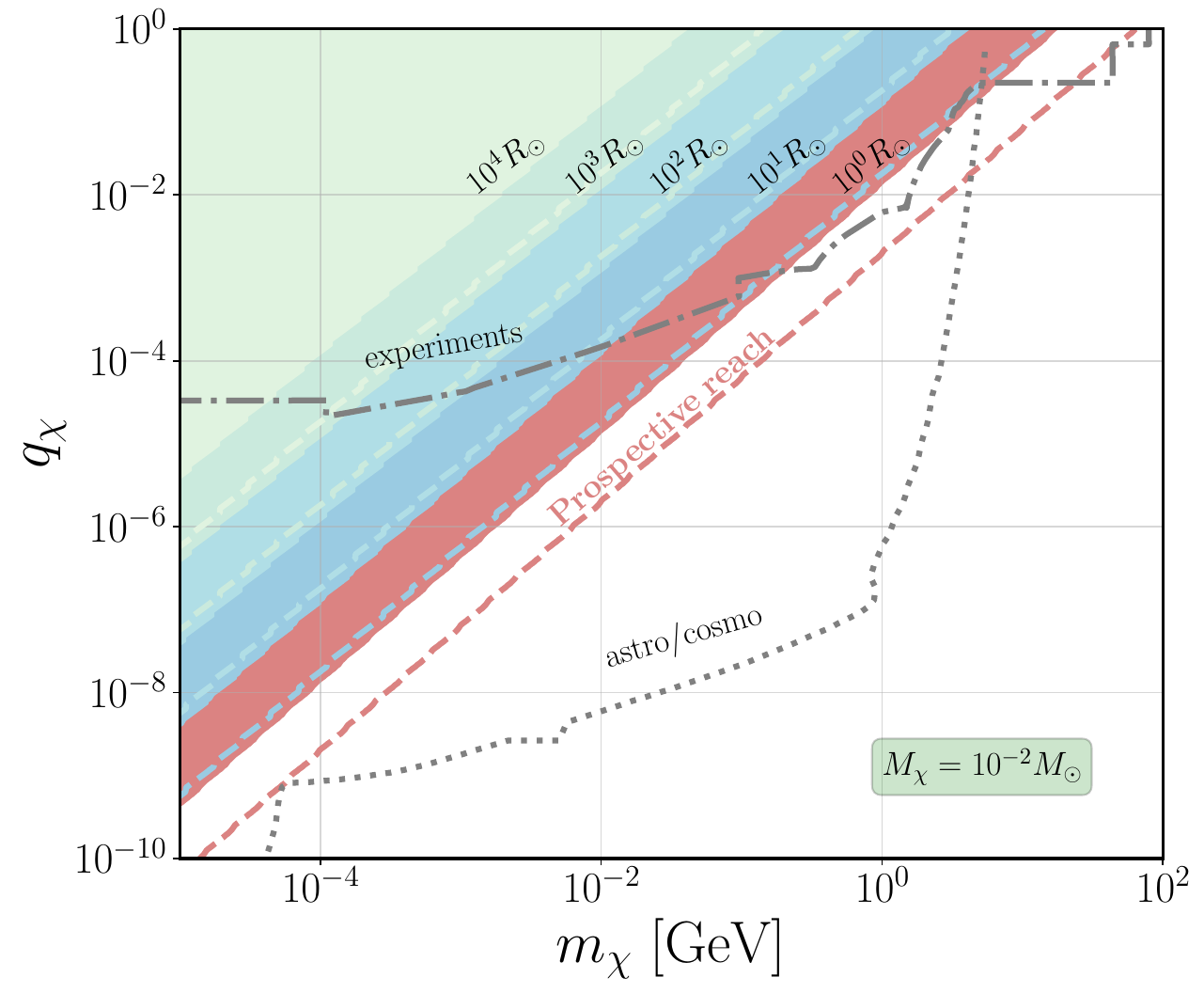}
\caption{\label{fig:kin_mix_consts_future} Similar to Fig. \ref{fig:consts_sigma}, but for the effective fractional charge $q_\chi$ for millicharged DM, assuming a minimum event time of $t_{E,\mathrm{min}} = 0.01$ days. The shaded regions assume a dimming fraction of $\mu_0 = 0.34$ while the dashed lines assume a lower dimming of $\mu_0 = 0.01$ for the same color. We have indicated in red the constraints for a prospective reach, where due to the smaller minimum event time we are more sensitive to smaller objects. All other shaded regions remain unchanged in the case with $t_{E, \mathrm{min}} = 1$ day. Only $m_\chi$ and $q_\chi$ are varied between the curves, with $\alpha_D = 0.1$. Overlaid dotted lines labelled astro/cosmo correspond to a combination of constraints from SN1987A \cite{Chang:2018rso, Fiorillo:2024upk}, stellar evolution \cite{Vogel:2013raa, Fung:2023euv}, and limits on $\Delta N_\mathrm{eff}$ from BBN and CMB \cite{Vogel:2013raa, Adshead:2022ovo}. Overlaid dotted-dashed lines correspond to constraints from a variety of experiments, including colliders \cite{Davidson:2000hf}, SLAC \cite{Prinz:1998ua}, OPOS \cite{Badertscher:2006fm}, ArgoNeuT \cite{ArgoNeuT:2019ckq}, BEBC \cite{BEBC} and milliQan \cite{milliQan:2021lne}.}
\end{figure}

While we have assumed that parameters in the event rate expressions Eqs. \eqref{eq:differential_events} and \eqref{eq:N_events} for dimming events were similar to those of microlensing, different parameters can lead to higher sensitivities (and therefore stronger constraints). In particular, while the minimum event time for microlensing in OGLE-III+IV was $t_{E,\mathrm{min}} = 1$ day \cite{Mroz:2024mse, Mroz:2024wag}, if the efficiency for dimming effects was permitted to be non-zero at lower event times, then the surveys would be sensitive to smaller sizes. Some microlensing surveys, such as Subaru/HSC \cite{Niikura:2017zjd}, and upcoming surveys such as the Legacy Survey of Space and Time (LSST) at the Vera Rubin Observatory \cite{LSST} and the Roman telescope \cite{Roman}, have already considered using much lower event times.  Existing analyses also consider reduced minimum event time of 2 minutes \cite{Croon:2020ouk} and 15 minutes \cite{DeRocco:2023gde, DeRocco:2023hij, Fardeen2024}, respectively. Additionally, if the threshold for a minimum amount of dimming was lower rather than $\mu_0=0.34$ (chosen to mirror the amplification factor for microlensing), then the surveys could be more sensitive to more dilute DM clumps. A lower threshold for dimming is feasible since OGLE is sensitive to variations in brightness at the milli-magnitude level \cite{ogle-iv}.

To demonstrate these effects, Fig. \ref{fig:kin_mix_consts_future} shows limits on $m_\chi$ and $q_\chi$ assuming a lower minimum event time of $t_{E, \mathrm{min}} = 0.01$ days for two different dimming thresholds. The shaded areas correspond to a dimming of $\mu_0 = 0.34$, while the dashed lines correspond to $\mu_0 = 0.01$. Considering a shorter minimum event time does not change the reach of the constraints at clump sizes already observable with higher event times. Instead, a shorter minimum event time renders surveys sensitive to smaller clumps which otherwise transit a star too quickly to be seen. Clumps with physical sizes $R_\chi \geq 10 R_\odot$ are already detectable at $t_{E,\mathrm{min}} = 1 $ day (indicated in blue/green shaded areas), whereas clumps with sizes $R_\chi = R_\odot$ were not detectable at the higher minimum event time. 

There are also constraints on millicharged DM in this region of parameter space from interstellar gas cloud cooling \cite{Bhoonah:2018wmw,Bhoonah:2018gjb,Bhoonah:2020dzs,Wadekar:2019mpc,Wadekar:2022ymq}, which will depend non-trivially on the modeling of millicharged clumps heating local portions of gas clouds. Additionally, we have not shown the constraint obtained by requiring that the DM-baryon fluid at recombination is completely decoupled \cite{McDermott:2010pa}, as this is a stronger cosmological condition than necessary.

\textbf{\textit{Conclusion.}}---In this Letter, we have shown how existing microlensing surveys can observe dimming of distant stars due to transits of clouds of DM that interact with photons. We have shown that microlensing surveys are more sensitive to dimming effects than microlensing events for DM clumps ($c.f.$~\cite{Croon:2020ouk}) larger than the Einstein radius of the object. The null observation of dimming effects in microlensing surveys such like EROS-2 and OGLE-III+IV could be used to limit the size, mass, and abundance of DM clumps, as well as limit the microphysical properties of DM comprising the clumps, such as the mass, photon interaction cross sections, and the effective charge of the DM particle in a millicharged DM scenario.  Searching existing survey data for dimming events, which will require no modifications to existing or upcoming microlensing data, would make the constraints suggested in this letter concrete and is an obvious next step.  These searches are complementary or competitive with existing DM-photon interaction searches, depending on the survey observing cadence and sensitivity to changes in source brightness. Further, with dedicated modeling of dimming events, any observed dimming could give insight into the properties and abundance of dark clumps and the DM that composes them. We emphasize that the majority of our analysis was performed model-independently. This approach could be used to look for or constrain any dark sector model that interacts with photons and forms compact objects.

We note several additional possible extensions to our treatment. First, we have not considered how the lampshade effect and microlensing might overlap. DM clumps with densities in an intersecting region of sensitivity ($c.f.$ Fig. \ref{fig:heuristic_params}) can be both a lampshade and a gravitational lens. Here, the competing dimming and brightening effects could result in a lower number observed microlensing or dimming events. There are, however, additional signatures to look for, as the star may both brighten and dim during different points in the transit if microlensing and dimming effects dominate at different radii. Second, our analysis can be generalized by considering various density profiles for the dark clumps (see Refs. \cite{Croon:2020wpr, Croon:2020ouk} for the effects of density profiles in the case of microlensing) and considering other surveys such as Subaru/HSC \cite{Niikura:2017zjd}, as well as the future LSST \cite{LSST} and Roman Space Telescope \cite{Roman}. These surveys allow for source resolution (see $e.g.$ Refs.~\cite{Croon:2020ouk, Winch:2020cju, DeRocco:2023hij, DeRocco:2023gde}), lower minimum observation times, and better sensitivities to changes in brightness -- all of which would result in improved bounds. Finally, we note that considering light from an extended source should increase survey dimming event sensitivity.

\textbf{\textit{Acknowledgements.}}---We thank Gonzalo Alonso-\'Alvarez, Christopher V. Cappiello, Wendy Crumrine, David Curtin, Audrey Fung, Vera Gluscevic, Przemek Mr{\'o}z, Nirmal Raj, and Han Wu for useful discussions. The authors and this work were supported by the Arthur B. McDonald Canadian Astroparticle Physics Research Institute, the Natural Sciences and Engineering Research Council of Canada (NSERC), and the Canada Foundation for Innovation. Research at Perimeter Institute is supported by the Government of Canada through the Department of Innovation, Science, and Economic Development, and by the Province of Ontario. JLK is supported by an NSERC Canada Graduate Scholarship - Doctoral (CGS-D).

\bibliography{refs}

\newpage\hbox{}\thispagestyle{empty}\newpage

\title{Supplementary Material: Dimming Starlight with Dark Compact Objects}
\maketitle
\setcounter{equation}{0}
\setcounter{figure}{0}
\setcounter{table}{0}
\setcounter{page}{1}
\makeatletter
\renewcommand{\theequation}{S\arabic{equation}}
\renewcommand{\thefigure}{S\arabic{figure}}
\setcounter{section}{0}
\renewcommand{\thesection}{S-\Roman{section}}

\title{Supplementary Material: Dimming Starlight with Dark Compact Objects}

\section{Effective radius}

\begin{figure}
\includegraphics[width=\columnwidth]{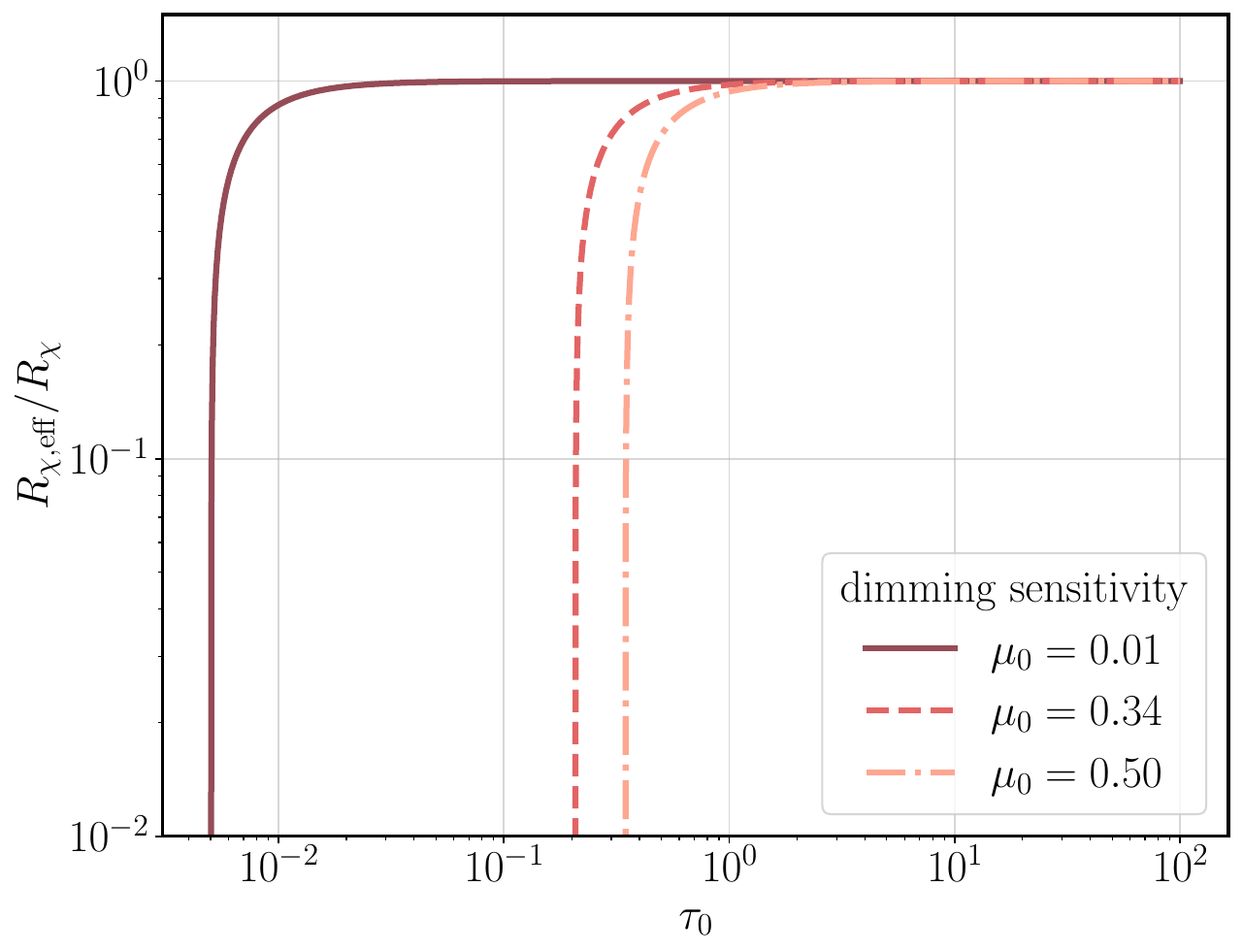}
\caption{\label{fig:R_X_eff} The effective radius $R_{\chi, \mathrm{eff}}$ at which the source star is dimmed by a fraction $\mu_0$ as a function of the characteristic optical depth $\tau_0$. The solid, dashed, and dotted-dashed curves correspond to effective radii assuming $\mu_0 = 0.01, 0.34$, and $0.50$ respectively.}
\end{figure}

In the main text, we provide an expression for the effective radius, which we again write for convenience,
\begin{align}
    R_{\chi,\mathrm{eff}} = R_\chi \sqrt{1 - \frac{1}{4\tau_0^2} [\ln (1 - \mu_0)]^2  . \label{eq:R_eff}
}
\end{align}
The relationship between the effective radius of the cloud in terms of its physical radius, and the cloud's characteristic optical depth is shown in Fig. \ref{fig:R_X_eff}, which shows the effective radius as a function of optical depth for three different values of $\mu_0$. For the fiducial amount of dimming $\mu_0 = 0.34$ considered throughout the main text, a characteristic optical depth of $\tau_0 = 1$ would result in $R_{\chi, \mathrm{eff}} \approx R_\chi$.

\section{Event rates}

\begin{figure}
\includegraphics[width=\columnwidth]{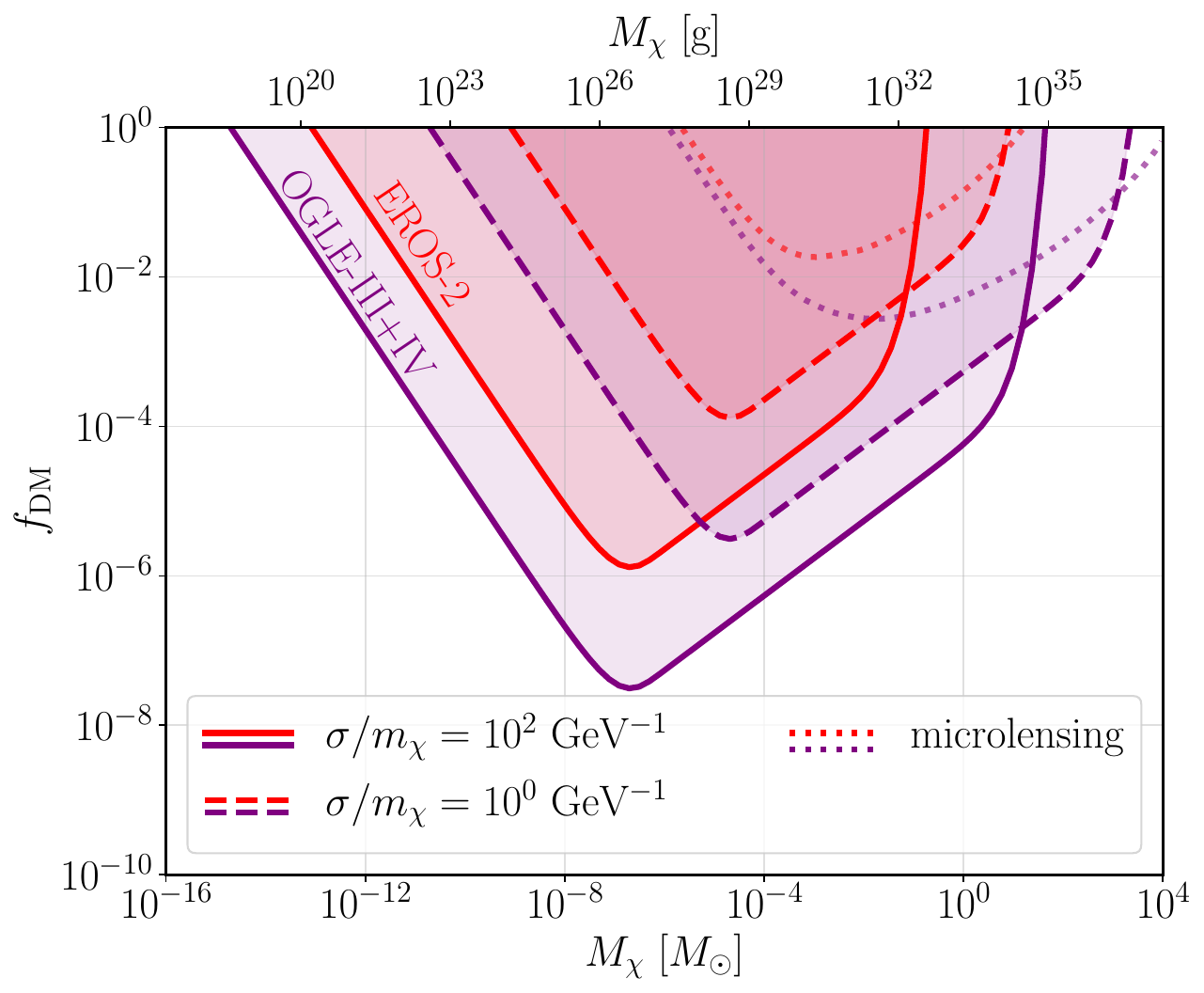}
\caption{\label{fig:f_example_ogle} Limits on the fraction of dark matter $f_\mathrm{DM}$ inside clumps of mass $M_\chi$ assuming null observations of dimming effects in microlensing surveys, assuming a fixed characteristic optical depth of $\tau_0 =1$. The red (purple) shaded areas correspond to restrictions on $f_\mathrm{DM}$ from the EROS-2 LMC (OGLE-III+IV) survey, with sizes $R_\chi$ computed at fixed values of $\sigma / m_\chi$. The dotted curves in red and purple show restrictions on $f_\mathrm{DM}$ assuming microlensing from point sources from EROS-2 LMC \cite{Croon:2020wpr} and OGLE-III+IV \cite{Mroz:2024mse, Mroz:2024wag}, respectively.}
\end{figure}

Appendix B of Ref.~\cite{Croon:2020wpr} has shown that the integral in the expected number of dimming/microlensing events can be efficiently computed, since the integral over $t_E$ has an analytical solution. 
In this work we evaluate the integral according to
\begin{align}
    A \int dt_E \frac{e^{-B/t_E^2}}{t_E^4} = \frac{A}{2B} \left[ \frac{e^{-B/t_E^2}}{t_E} - \frac{\sqrt{\pi}}{2} \frac{\text{erf} (\sqrt{B}/t_E)}{\sqrt{B}} \right],
\end{align}
which we include here for completeness.

Note that in this work we have only considered dimming events in the LMC due to DM clumps in the MW. However, there are also events expected from clumps in the LMC, which were considered in Refs.~\cite{Mroz:2024mse}. We find that events due to clumps from the LMC were always subdominant to those from the MW, improving the number of events at most by a factor of two, which agrees with the results found in Ref.~\cite{Mroz:2024mse}, in which the number of events were dominated by MW objects except at small clump masses. For simplicity, we do not consider the additional events from clumps in the LMC. However, considering clumps in the LMC would only strengthen the sensitivity/constraints shown, as they imply more events.

Fig. \ref{fig:f_example_ogle} shows limits on the fraction of dark matter $f_\mathrm{DM}$ that could reside in dark matter clumps. Similar to Fig. (2) in the main text, to compute these limits on $f_\mathrm{DM}$, we have taken $N_\mathrm{thresh} = 3.9$ events as the threshold number of events. In more detail, Fig. \ref{fig:f_example_ogle} shows constraints on $f_\mathrm{DM}$ for two different values of $\sigma/m_\chi$, for a fixed characteristic optical depth of $\tau_0 = 1$ so that $R_{\chi, \mathrm{eff}} \approx R_\chi$, where $R_\chi$. We have shown both EROS-2 LMC constraints in red and OGLE-III+IV LMC constraints in purple for comparison. We have also plotted constraints on $f_\mathrm{DM}$ from point-like lenses in EROS-2 \cite{Croon:2020ouk} and OGLE-III+IV \cite{Mroz:2024mse, Mroz:2024wag} in dotted curves. Having a longer total observation time and a larger number of source stars allows one to limit $f_\mathrm{DM}$ to lower values, while having a longer maximum event time $t_{E, \mathrm{max}}$ allows for access to higher clump masses due to the $R_\chi \propto M_\chi^{1/2}$ relationship. 

Note that in Fig. \ref{fig:f_example_ogle} we assume that there would be no candidate dimming events in the indicated surveys, for a light dimming of a factor $\mu_0 = 0.34$ (where the brightening event threshold in microlensing surveys is similarly 0.34). We stress that the dimming analysis has not yet been undertaken. It will require a foreground analysis of the expected number of dimming events from non-DM astrophysical events. In this work we assume that no candidate dimming events have been found in the data for our estimates, and leave a more careful study of the individual light curves in the data for a future study. In addition, while we have overlaid microlensing constraints from point lenses in Fig. \ref{fig:f_example_ogle}, we comment that for extended DM objects which are large enough, the constraints on $f_\mathrm{DM}$ can vary drastically depending on the assumed density profile of the object. In general, microlensing constraints get weaker as the object becomes less dense \cite{Croon:2020wpr}. Therefore to compare the possible constraints from dimming, we have only shown the microlensing constraints from point-like lenses.

\section{DM-SM photon scattering}

\begin{figure*}
\includegraphics[width=\textwidth]{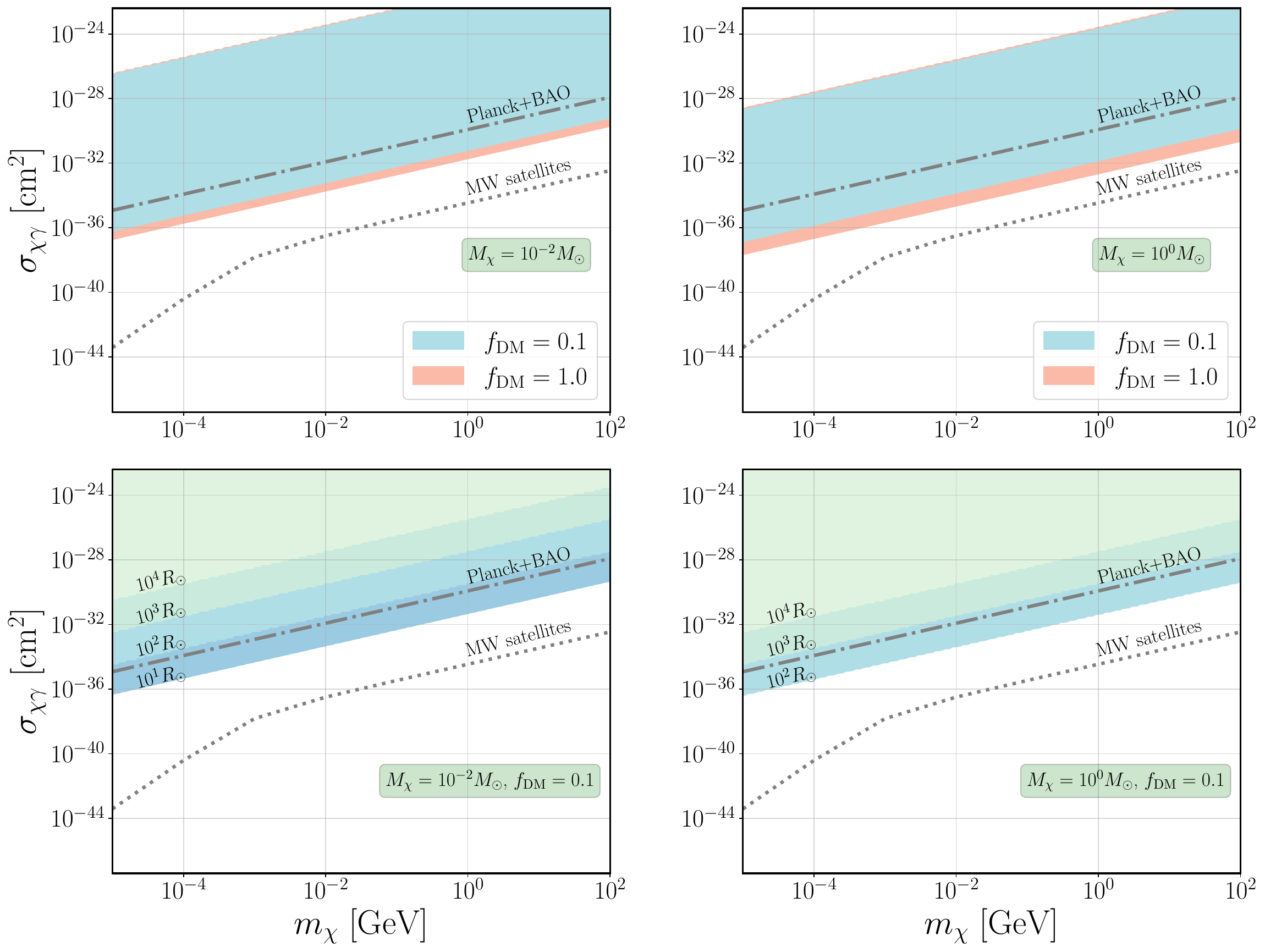}
\caption{\label{fig:consts_comb} Constraints that could be placed on $m_\chi$ and $\sigma_{\chi \gamma}$ for elastic scattering of SM photons and $\chi$-particles from a future null analysis of dimming events in the OGLE-III+IV surveys. \textbf{Top panels:} Constraints at a fixed characteristic depth $\tau_0=1$, where the orange (blue) shaded regions are constraints, assuming that $f_\mathrm{DM} = 1$ ($f_\mathrm{DM} = 0.1$) of DM is inside of clumps of masses $M_\chi = 10^{-2} M_\odot$ in the left panel and $M_\chi = 10^{0} M_\odot$ in the right panel. \textbf{Bottom panels}: Constraints at varying optical depths and fixed physical sizes $R_\chi$ from $10$ to $10^4 R_\odot$. In all panels, the overlaid gray dashed-dotted line are bounds on the elastic scattering cross section from Planck+BAO \cite{Becker:2020hzj}, while the gray dotted line are bounds from MW satellite galaxies \cite{Crumrine:2024sdn}. We note the MW satellite constraints are calculated for $f_{\text{DM}}=1$ and for dark matter that has not formed structures such as those considered in this work, causing them to likely be quite different for the scenarios covered in this work.}
\end{figure*}

We show in Fig. \ref{fig:consts_comb} a 4-panel figure which shows constraints that can be placed on the dark matter mass $m_\chi$ and the DM-photon elastic scattering cross section $\sigma_{\chi \gamma}$. The top panels correspond to a fixed optical depth of $\tau_0= 1$ which in turn implies an effective radius of $R_{\chi, \mathrm{eff}} \approx R_\chi$. We show in orange (blue) constraints assuming $f_\mathrm{DM} = 1$ ($f_\mathrm{DM} = 0.1$) fraction of DM in compact objects at fixed masses of $M_\chi = 10^{-2} M_\odot$ in the left panel and $M_\chi = 10^0 M_\odot$ in the right panel. The constraints fall off at low values of dark matter mass $m_\chi$ and high values of $\sigma_{\chi \gamma}$ because these would result in smaller DM clump sizes (for fixed optical depth), which would avoid detection in the microlensing surveys as they are limited by the minimum event time $t_{E,\mathrm{min}}$. On the other side of the sensitivity band, for lower values of $m_\chi$ and higher values of $\sigma_{\chi \gamma}$, the constraints get weaker since the sizes of the objects would be too large and are therefore limited by the maximum event time $t_{E, \mathrm{max}}$. 

The bottom panels of Fig. \ref{fig:consts_comb} are similar to Fig. (3) in the main text, for two different clump masses $M_\chi = 10^{-2} M_\odot$ (left) and $M_\chi = M_\odot$ (right). Note that the bottom-left panel of Fig. \ref{fig:consts_comb} is Fig. (3) in the main text. See main text for more details.

\section{Millicharged dark matter}

\begin{figure*}
\includegraphics[width=\textwidth]{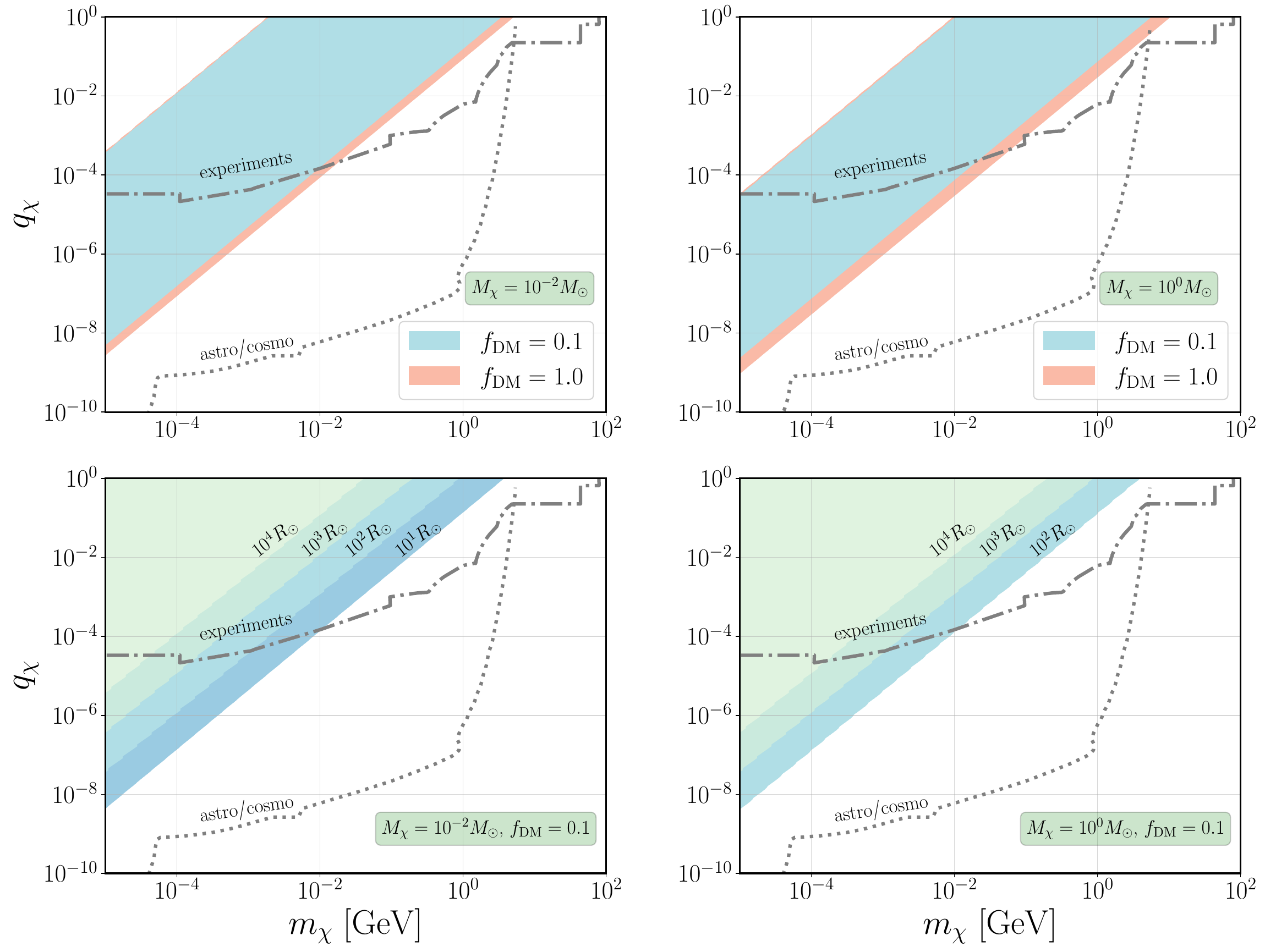}
\caption{\label{fig:kin_mix_consts_comb} Constraints on the millicharged particle mass $m_\chi$ and effective fractional charge $q_\chi$ from null observations of dimming events in OGLE-III+IV, similar to Fig. \ref{fig:consts_comb}. \textbf{Top panels:} Constraints at varying physical clump sizes and a fixed characteristic depth $\tau_0=1$, where the orange (blue) shaded regions are constraints, assuming that $f_\mathrm{DM} = 1$ ($f_\mathrm{DM} = 0.01$) of DM is inside of clumps of masses $M_\chi = 10^{-2} M_\odot$ in the left panel and $M_\chi = 10^{0} M_\odot$ in the right panel. \textbf{Bottom panels}: Constraints at varying optical depths and fixed physical sizes $R_\chi$ from $10$ to $10^4 R_\odot$. Overlaid dotted lines labelled astro/cosmo correspond to a combination of constraints from SN1987A \cite{Chang:2018rso, Fiorillo:2024upk}, stellar evolution \cite{Vogel:2013raa, Fung:2023euv}, and limits on $\Delta N_\mathrm{eff}$ from BBN and CMB \cite{Vogel:2013raa, Adshead:2022ovo}. Overlaid dotted-dashed lines correspond to constraints from a variety of experiments, including colliders \cite{Davidson:2000hf}, SLAC \cite{Prinz:1998ua}, OPOS \cite{Badertscher:2006fm}, ArgoNeuT \cite{ArgoNeuT:2019ckq}, BEBC \cite{BEBC} and milliQan \cite{milliQan:2021lne}. There are also constraints on millicharged DM from interstellar gas cloud cooling \cite{Bhoonah:2018wmw,Bhoonah:2018gjb,Bhoonah:2020dzs,Wadekar:2019mpc,Wadekar:2022ymq}, which will depend on future modeling of millicharged clumps heating local portions of gas clouds.}
\end{figure*}

Fig. \ref{fig:kin_mix_consts_comb} shows constraints that could be placed on $m_\chi$ and $q_\chi$ from the null observation of DM dimming events from OGLE-III+IV, assuming $\alpha_D = 0.1$. All panels and their behaviour are similar to Fig. \ref{fig:consts_comb}, described above. We have also overlaid constraints on millicharged particles from astrophysical/cosmological considerations such as SN1987A \cite{Chang:2018rso, Fiorillo:2024upk}, stellar evolution of horizontal branch stars, red giants, and white dwarfs \cite{Vogel:2013raa, Fung:2023euv}, and limits on $\Delta N_\mathrm{eff}$ from big bang nucleosynthesis (BBN) and the cosmic microwave background (CMB) \cite{Vogel:2013raa, Adshead:2022ovo}, as well as from experiments, comprised of a combination of constraints from colliders \cite{Davidson:2000hf}, SLAC \cite{Prinz:1998ua}, OPOS \cite{Badertscher:2006fm}, ArgoNeuT \cite{ArgoNeuT:2019ckq}, BEBC \cite{BEBC} and milliQan \cite{milliQan:2021lne}. 

As discussed in the main text, comparing between the bottom-left panel of Fig. \ref{fig:kin_mix_consts_comb} and Fig. (4) in the main text, we see that considering a shorter minimum event time does not change the reach of the constraints at clump sizes which would already have created observable dimming for a higher minimum event time.

On the other hand, decreasing the required amount of dimming from  $\mu_0 =0.34$ to $\mu_0 = 0.01$ is advantageous because for a given clump size, dimming events from clumps are observable at higher values of the DM mass $m_\chi$ and lower values of the charge $q_\chi$. This is because the required optical depth of the clump, $\tau_{0, \mathrm{min}}$, is smaller for a lower dimming threshold. We reiterate that a threshold of $\mu_0 = 0.01$ is obtainable, for instance, in OGLE-IV as it can be sensitive to changes in brightness at the milli-magnitude level for the brightest stars \cite{ogle-iv}.

Hence the strongest bounds arise when we consider both a lower minimum event time and a lower threshold for a dimming event as in Fig. (4) of the main text, which yields sensitivity to the smallest clump sizes and largest effective radii. For the case of solar-radius-size clumps, with sizes $R_\chi = 10^0 R_\odot$ with a minimum event time of $t_{E, \mathrm{min}} = 0.01$ days and a dimming threshold of $\mu_0 = 0.01$, we see null observations of dimming events would compete with other bounds for millicharged particles at both $m_\chi \approx 10^{-4}$ GeV and $m_\chi \approx 10$ GeV.

\end{document}